\documentclass[aps, prd, twocolumn, lengthcheck, superscriptaddress, 
nofootinbib]{revtex4-1}

\usepackage{amsmath,amssymb,amsfonts,physics,graphicx,float}
\graphicspath{{pic/}}
\usepackage[setpagesize=false]{hyperref}
\allowdisplaybreaks[4]

\usepackage[labelformat=simple]{subcaption} 

\usepackage{ulem}
\usepackage{color}

\def\lsim{\mathrel{\rlap{\lower3pt\hbox{\hskip1pt$\sim$}}
		\raise1pt\hbox{$<$}}} 
\def\gsim{\mathrel{\rlap{\lower3pt\hbox{\hskip1pt$\sim$}}
		\raise1pt\hbox{$>$}}} 

\definecolor{ar}{rgb}{1.0, 0.01, 0.24}
\definecolor{al}{rgb}{0.82, 0.1, 0.26}
\definecolor{ev}{rgb}{0.56, 0.0, 1.0}

\def\be{\begin{eqnarray}}
\def\ee{\end{eqnarray}}

\begin{document}

\title{
Exploring the first-order phase transition in neutron stars using the parity doublet model and NJL-type quark model
}

\author{Bikai Gao}
\email{gaobikai@hken.phys.nagoya-u.ac.jp}
\affiliation{Department of Physics, Nagoya University, Nagoya 464-8602, Japan}
\affiliation{School of Frontier Sciences, Nanjing University, Suzhou, 215163, China}

\author{Wen-Li Yuan}
\email{wlyuan@pku.edu.cn}
\affiliation{School of Physics and State Key Laboratory of Nuclear Physics and Technology, Peking University, Beijing 100871, China}

\author{Masayasu Harada}
\email{harada@hken.phys.nagoya-u.ac.jp}
\affiliation{Department of Physics, Nagoya University, Nagoya 464-8602, Japan}
\affiliation{Kobayashi-Maskawa Institute for the Origin of Particles and the Universe, Nagoya University, Nagoya, 464-8602, Japan}

\author{Yong-Liang Ma}
\email{ylma@nju.edu.cn}
\affiliation{School of Frontier Sciences, Nanjing University, Suzhou, 215163, China}

\date{\today}

\begin{abstract}
We investigate the possibility and impacts of a first-order phase transition from hadronic matter to quark matter in neutron stars (NSs) using two specific models: the parity doublet model (PDM) for the hadronic phase and the Nambu-Jona-Lasinio (NJL) type model for the quark phase. By combining these models, we construct hybrid equations of state (EOSs) that capture the transition between the two phases. We explore the parameter space of both models to identify the conditions under which a first-order phase transition can occur and study its effects on NS properties. We identify the suitable parameter space and constrain the onset density of the first-order phase transition. For $m_0$ = 500 MeV---the chiral invariant mass in PDM, the phase transition occurs between 1.9$n_0$ and 2.95$n_0$ and ends between 2.1$n_0$ and 3.6$n_0$. Increasing $m_0$ to 600 MeV shifts the phase transition to higher densities, occurring between 2.9$n_0$ and 4.1$n_0$ and ending between 3.4$n_0$ and 4.6$n_0$.


\end{abstract}

\maketitle


\section{Introduction}
Quantum Chromodynamics (QCD) is believed to be the fundamental theory that describes the strong interactions between quarks and gluons. Under normal conditions, quarks and gluons are confined within hadrons, such as protons, neutrons, and pions. These composite particles serve as effective degrees of freedom (d.o.f.) for describing the physics in vacuum and low density/temperature. However, the situation changes with the increasing of temperature or the density as hadrons begin to overlap and quarks and gluons d.o.f. will come into play. Such extreme forms of matter are believed to have existed in the early universe and are present in astrophysical objects like neutron stars (NSs) (see, e.g., Refs.~\cite{Fukushima:2010bq,Fukushima:2013rx,Baym:2017whm,Ma:2019ery,Annala:2019puf,Lovato:2022vgq,Brandes:2023bob} and references therein). 

Along the temperature axis in hot QCD, a heated hadron resonance gas smoothly transforms into a quark-gluon plasma, despite the apparent differences between these two phases. Experiments have successfully created such hot matter through high-energy heavy ion collisions, where the colliding energy is converted into heat. A coherent picture of hot QCD matter has emerged through the combined efforts of experimental analyses, ab-initio lattice simulations of QCD, and model calculations~\cite{Fukushima:2010bq,Lovato:2022vgq,Fukushima:2011jc,Petreczky:2012rq}. 

However, the situation is much different for cold dense matter, especially about the exact nature of the transition from nuclear matter to quark matter. Theoretically, this cannot be accessed due to the lack of solid approach anchored on nonperturbative QCD. Moreover, the model dependent predictions cannot be diagnosed by terrestrial experiments since such high density cannot be produced or Lattice QCD simulation due to the sign problem. The only constraints on nuclear matter properties are from the experimental data around the saturation density $n_0$ ($n_0 = 0.16$ fm$^{3}$) and predictions from the perturbative QCD at extreme high density $\gsim 50 n_0$.

Since the detection of the gravitational waves (GWs) from a binary NS merger GW170817~\cite{PhysRevLett.119.161101,LIGOScientific:2017ync,LIGOScientific:2018cki}, astrophysics enters the multi-messenger era. In combination with the observation of massive pulsars around two-solar mass~\cite{Miller:2021qha, Riley:2021pdl, Vinciguerra:2023qxq}, NS has become a natural laboratory for investigating the properties of matter under extreme conditions, particularly at ultra-high densities up to $\sim 10 n_0$. The characteristics of neutron stars are determined by the equation of state (EOS) of the matter composing them. Therefore, the theoretically obtained EOSs can be constrained by observational data, such as the mass-radius ($M\text{-}R$) relation of a NS, the tidal deformability, and the signals of gravitational waves stemming from binary NS mergers~\cite{Annala:2019puf,Brandes:2023bob,Hotokezaka:2011dh,Hotokezaka:2013iia,DePietri:2015lya,Maione:2016zqz,Moustakidis:2016sab,Radice:2016rys,Paschalidis:2017qmb,Annala:2017llu,Torres-Rivas:2018svp,Most:2018eaw,Tews:2018iwm,Tsang:2018kqj,Bauswein:2018bma,Orsaria:2019ftf,Most:2019onn,Yang:2020ucv,Sekiguchi:2011mc,Choi:2020eun,Annala:2021gom,Raaijmakers:2021uju,Prakash:2021wpz,Prakash:2021wpz,Kedia:2022nns,Fujimoto:2022xhv,Guo:2023som}.

One of the key questions in the study of NS EOSs is whether there is a possible transition from hadronic matter to quark matter---either a phase transition or a crossover. Such a transition could have significant implications for the structure and properties of NSs (see, e.g.,~\cite{Baym:2017whm, Lenzi:2012xz, Kojo:2020krb,Benic:2014jia}). When the transition is a first-order phase transition, there can be a sudden release or absorption of energy, leading to changes in the equilibrium state of the star. This can manifest as a sudden change in the radius of the neutron star, as the internal pressure and composition adjust to the new equilibrium state.

In previous works, by assuming the transition is a cossover transition, some of us constructed unified EOSs by interpolating the hadronic EOS from a hadronic model
based on the parity doublet structure and the quark EOS from an  Nambu-Jona-Lasinio (NJL)-type quark model~\cite{Minamikawa:2020jfj,Minamikawa:2021fln,Gao:2022klm,Kong:2023nue,Gao:2024chh}. In this paper, we devote ourselves to investigate the possibility and the impacts of the first-order phase transition in dense matter using the parity doublet model (PDM) for the hadronic phase and the NJL-type model for the quark phase. 

The PDM offers a unique perspective of baryon structure by incorporating chiral symmetry and its restoration, allowing the parity partners to have a degenerate chiral invariant mass $m_0$ at high densities when chiral symmetry is restored. The notion of chiral invariant mass was first introduced in \cite{PhysRevD.39.2805,10.1143/PTP.106.873} and is consistent with the recent lattice QCD simulation \cite{Aarts:2015mma,Aarts:2017rrl,Aarts:2018glk} as well as the skyrmion crystal approach to nuclear physics~\cite{Ma:2013ooa,Ma:2013ela}. Effective models based on parity doublet structure is widely used for studying the hadron structure and nuclear/NS physics\cite{Nishihara:2015fka,Minamikawa:2023ypn,HATSUDA198911,PhysRevC.75.055202,PhysRevC.77.025803,PhysRevC.82.035204,PhysRevD.84.034011,GALLAS201113,PhysRevC.84.045208,PhysRevD.85.054022,PhysRevC.87.015804,PhysRevD.88.105019,PhysRevC.96.025205,PhysRevC.97.045203,PhysRevC.97.065202,refId01,universe5080180,Motohiro,PhysRevC.100.025205,Mukherjee:2017jzi,Gao:2022klm,Marczenko:2022hyt,Minamikawa:2023eky,Baym:2017whm,Minamikawa:2020jfj,Minamikawa:2021fln,Marczenko:2019trv,Kong:2023nue, Gao:2024mew} . The key feature of the PDM is that a larger $m_0$ leads to a weaker $\sigma$ coupling to nucleon since nucleon does not have to acquire its mass entirely from the $\sigma$ field. The corresponding $\omega$ field also becomes smaller due to the equilibrium state at the saturation density $n_0$. Then at density larger than $n_0$, the $\sigma$ field decreases while the $\omega$ field increases. As a result, a larger $m_0$ would lead to a softer EOS.

By combining the PDM and the NJL-type model, we aim to construct a hybrid EOS that captures the transition from hadronic matter to quark matter. We will explore the parameter space of both models to identify the conditions under which a first-order phase transition can occur. The resulting EOSs will be compared with observational constraints from NS observation to assess their compatibility with the latest astronomical data.

This paper is organized as follows. In Sec.~\ref{sec-eos}, we present the EOSs used in our study. We describe the hadronic matter EOS based on the PDM and the quark matter EOS derived from the NJL-type model.
We also discuss the construction of the hybrid EOS incorporating a first-order phase transition.
In Sec.\ref{sec-NS}, we investigate the neutron star properties using the EOSs constructed in Sec.\ref{sec-eos}, focusing on the mass-radius relation and the constraints on the onset density of the phase transition.
Finally, we summarize our findings and discuss their implications in Sec.\ref{sec-conclusion}.


\section{EQUATION OF STATE }
\label{sec-eos}

In this section, we develop EOSs for neutron star matter in the hadron and quark phases separately, and then explore the unified EOS with a first-order phase transition by integrating them.

\subsection{NUCLEAR MATTER EOS}
\label{sec:PDM matter}

Through out this work, we assume that the hyperons do not enter into the matter in hadronic phase  and consider the case $N_f =2$. Following Ref.~\cite{Motohiro}, we express the thermodynamic potential based on parity doublet structure as
\begin{equation}
\begin{aligned}
\Omega_{\mathrm{PDM}}= & V\left(\sigma\right)-V\left(\sigma_0\right)-\frac{1}{2} m_\omega^2 \omega^2-\frac{1}{2} m_\rho^2 \rho^2 \\
& -\lambda_{\omega \rho}\left(g_\omega \omega\right)^2\left(g_\rho \rho\right)^2 \\
& -2 \sum_{i=+,-} \sum_{\alpha=p, n} \int^{k_f} \frac{\mathrm{d}^3 \mathbf{p}}{(2 \pi)^3}\left(\mu_\alpha^*-E_{\mathrm{p}}^i\right).
\end{aligned}
\end{equation}
Here $i = +, -$ denote for the parity of nucleon---since in PDM, not only the positive parity baryons, proton and neutron, but also the negative parity baryons which are regarded as the parity partner of positive parity baryons are involved---and $E_{{\bf p}}^{i} = \sqrt{{\bf p}^{2} + m_{i}^{2}}$ is the energy of baryon with mass $m_{i}$ and momentum ${\bf p}$. The potential $V(\sigma)$ is taken as
\begin{align}
V(\sigma) = -\frac{1}{2}\bar{\mu}^{2}\sigma^{2} + \frac{1}{4}\lambda_4 \sigma^4 -\frac{1}{6}\lambda_6\sigma^6 - m_{\pi}^{2} f_{\pi}\sigma.
\end{align}

\begin{figure}[htbp]\centering
\includegraphics[width=0.9\hsize]{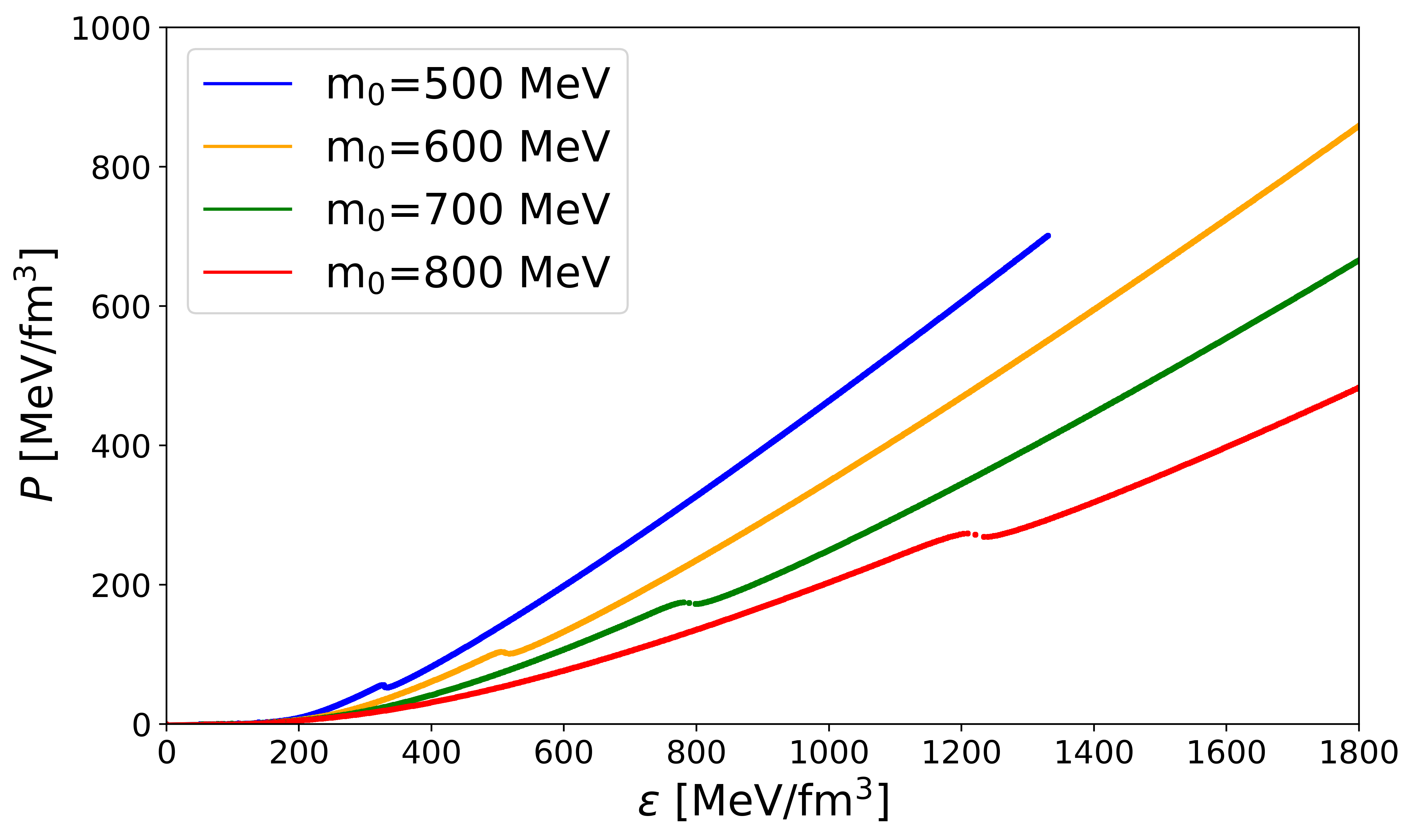}
\caption{EOS for different values of the chiral invariant mass $m_0$.}
\label{PDM_EOS}
\end{figure}

The total thermodynamic potential for the matter is obtained by including the effects of leptons as
\begin{align}
\Omega_{{\rm H}} = \Omega_{{\rm PDM}} + \sum_{l = e, \mu}\Omega_l,
\end{align}
where $\Omega_{l}(l=e,\mu)$ are the thermodynamic potentials for leptons,
\begin{equation}
\Omega_{l}=-2 \int^{k_{F}} \frac{d^{3} \mathbf{p}}{(2 \pi)^{3}}\left(\mu_{l}-E_{\mathbf{p}}^{l}\right).
\end{equation}
The parameters in the  PDM are determined by fitting the nuclear matter properties around saturation density $n_0$ and the pion decay constant for different $m_{0}$.  Using the explicit values evaluated in Ref. \cite{Gao:2024chh} we can then calculate the corresponding EOS shown in Fig.~\ref{PDM_EOS} in the hadronic phase with different choices of chiral invariant mass $m_0$.

\begin{figure*}[htbp]\centering
\includegraphics[width=0.9\hsize]{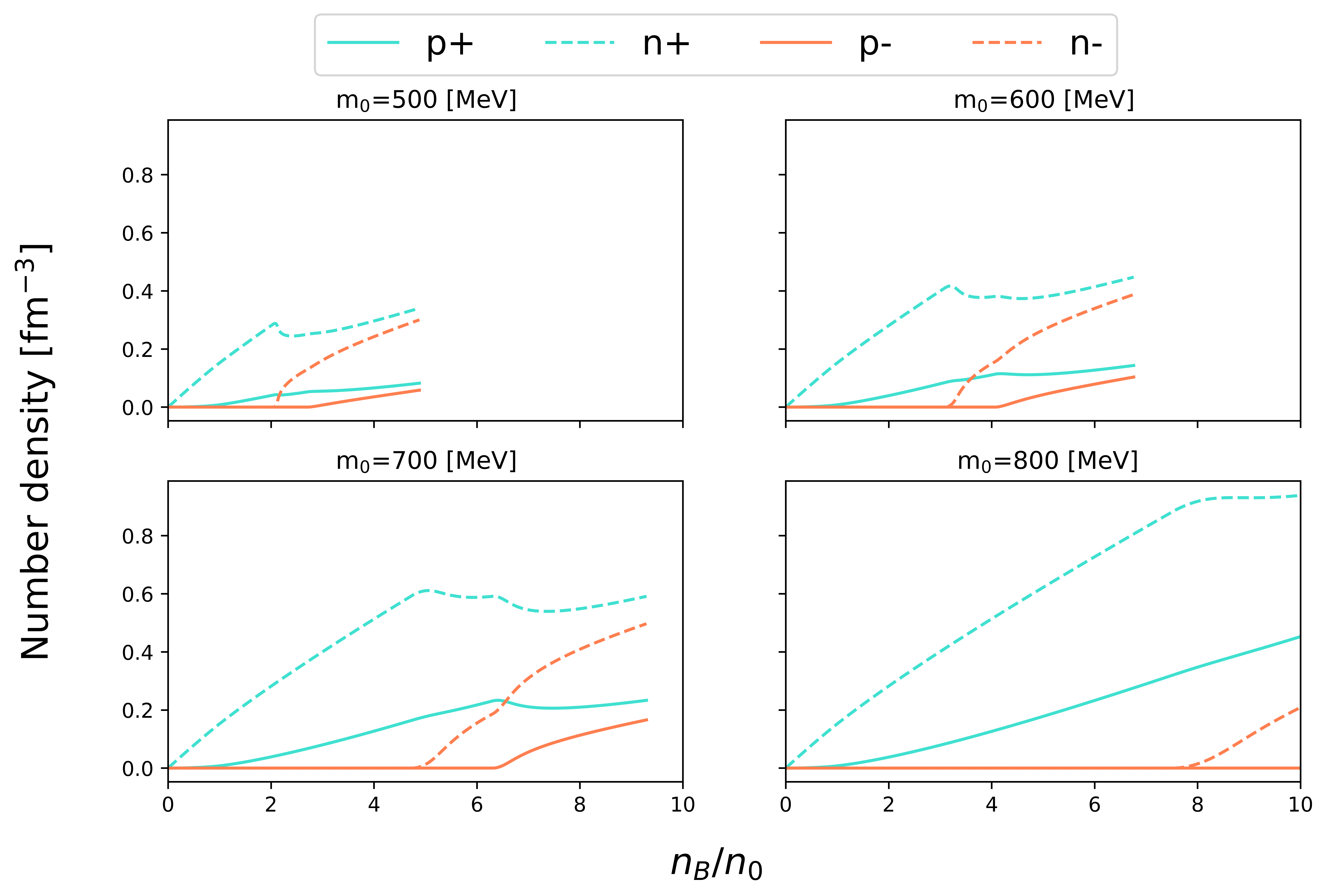}
\caption{Number density for different components with different values of the chiral invariant mass $m_0$. }
\label{Com_PDM}
\end{figure*}

From Fig. \ref{PDM_EOS}, one easily identify that a larger chiral invariant mass $m_0$---a feature of the PDM---leads to a relatively softer EOS. In addition, one can also observe a transition due to the entering of the nucleon with negative parity---another feature of the PDM, as shown in Fig.~\ref{Com_PDM}. In a subsequent section, we will employ the Maxwell construction to derive the EOS with a first-order phase transition within the PDM, which is associated with the inclusion of the negative parity nucleon.

The onset density of the  transition, characterized by the appearance of the negative parity nucleon N(1535), is strongly influenced by the choice of the chiral invariant mass $m_0$. In the PDM, the nucleon N(939) and its chiral partner N(1535) acquire their masses through two different mechanisms: the chiral invariant mass $m_0$, which is independent of the chiral condensate, and the mass generated by the chiral condensate, which is proportional to the sigma field $\sigma$. The mass difference between N(939) and N(1535) can be expressed as:
\be
\Delta m = g_1 \sigma + g_2 \sigma = (g_1 + g_2) \sigma
\ee
where $g_1$ and $g_2$ are coupling constants.
As the matter density increases, the chiral condensate $\sigma$ decreases due to the partial restoration of chiral symmetry. For a smaller value of $m_0$, the coupling constants $g_1$ and $g_2$ must be larger to reproduce the physical masses of N(939) and N(1535) in vacuum\footnote{The coupling constants $g_{1, 2}$ are calculated as 
\begin{align}
g_{1, 2} = \frac{1}{2f_{\pi}}\left(\sqrt{(m_{-} + m_{+})^{2} -4 m_0^2} \pm (m_- - m_+) \right). \nonumber
\end{align}
We can easily observe that smaller $m_0$ lead to larger value of $g_{1, 2}$.}. In that case, the mass difference $\Delta m$ is more sensitive to changes in the chiral condensate $\sigma$ when $m_0$ is small since the coupling constant $g_1$ and $g_2$ are comparatively larger. 
As a result, for a smaller $m_0$, the mass of N(1535) approaches the mass of N(939) more rapidly as the density increases. This means that N(1535) becomes energetically favorable and appears in the system at a lower density compared to the case with a larger $m_0$. This is confirmed in the explicit calculation shown in Fig. \ref{Com_PDM} that a smaller chiral invariant mass $m_0$ leads to an earlier  emergence of the negative parity baryon N(1535).

\subsection{QUARK MATTER EOS}
\label{NJL matter}
Following Refs.\cite{Baym:2017whm,Baym:2019iky}, we use the NJL-type model to describe the quark matter. 
The model includes three-flavors and U(1)$_A$ anomaly effects through the quark version of the Kabayashi-Maskawa-'t Hooft (KMT) interaction~\cite{Kobayashi:1970ji, tHooft:1986ooh}. 
The coupling constants are chosen to be the Hatsuda-Kunihiro parameters 
which successfully reproduce the hadron phenomenology at low energy \cite{Baym:2017whm, Hatsuda:1994pi}: 
$G\Lambda^{2}=1.835, K\Lambda^{5}=9.29$ with $\Lambda=631.4\, \rm{MeV}$, see the definition below.
The couplings $g_{V}$ and $H$ characterize the strength of the vector repulsion and attractive diquark correlations whose range will be examined later 
when we discuss the NS constraints.

We can then write down the thermodynamic potential as
\begin{equation}
\begin{aligned}
\Omega_{\mathrm{CSC}}
=&\, \Omega_{s}-\Omega_{s}\left[\sigma_{f}=\sigma_{f}^{0}, d_{j}=0, \mu_{q}=0\right] \\
&+\Omega_{c}-\Omega_{c}\left[\sigma_{f}=\sigma_{f}^{0}, d_{j}=0\right],
\end{aligned}
\end{equation}
where the subscript ``0" stands for the values in vacuum, and
\begin{align}
&\Omega_{s}=-2 \sum_{i=1}^{18} \int^{\Lambda} \frac{d^{3} \mathbf{p}}{(2 \pi)^{3}} \frac{\epsilon_{i}}{2} \label{energy eigenvalue},\\
&\Omega_{c}=\sum_{i=u,d,s}\left(2 G \sigma_{i}^{2}+H d_{i}^{2}\right)-4 K \sigma_{u} \sigma_{d} \sigma_{s}-g_{V} n_{q}^{2},
\end{align}
with $\sigma_{f}$ being the chiral condensates of quark flavor $f$, $d_{j}$ being quark condensates, and $n_{q}$ being the quark density. 
In Eq.(\ref{energy eigenvalue}), $\epsilon_{i}$ are energy eigenvalues obtained from inverse propagator in Nambu-Gorkov bases
\begin{equation}
S^{-1}(k)=\left(\begin{array}{lc}
\gamma_{\mu} k^{\mu}-\hat{M}+\gamma^{0} \hat{\mu} & \gamma_{5} \sum_{i} \Delta_{i} R_{i} \\
-\gamma_{5} \sum_{i} \Delta_{i}^{*} R_{i} & \gamma_{\mu} k^{\mu}-\hat{M}-\gamma^{0} \hat{\mu}
\end{array}\right),
\end{equation}
where
\begin{equation}
\begin{aligned}
&M_{i} =m_{i}-4 G \sigma_{i}+K\left|\epsilon_{i j k}\right| \sigma_{j} \sigma_{k}, \\
&\Delta_{i} =-2 H d_{i} ,\\
&\hat{\mu} =\mu_{q}-2 g_{V} n_{q}+\mu_{3} \lambda_{3}+\mu_{8} \lambda_{8}+\mu_{Q} Q,\\
&(R_{1}, R_{2}, R_{3})=(\tau_{7}\lambda_{7}, \tau_{5}\lambda_{5}, \tau_{2}\lambda_{2}).
\end{aligned}
\end{equation}
$S^{-1}(k)$ is $72\times72$ matrix in terms of the color,
flavor, spin, and Nambu-Gorkov basis, which has 72 eigenvalues. $M_{u,d,s}$ are the constituent masses of $u, d, s$ quarks and $\Delta_{1,2,3}$ are the gap energies. 
The $\mu_{3,8}$ are the color chemical potentials which will be tuned to achieve the color neutrality. 
The total thermodynamic potential including the effect of leptons is 
\begin{equation}
\Omega_{\mathrm{Q}}=\Omega_{\mathrm{CSC}}+\sum_{l=e, \mu} \Omega_{l}.
\end{equation}
The mean fields are determined from the gap equations,
\begin{equation}
0=\frac{\partial \Omega_{\mathrm{Q}}}{\partial \sigma_{i}}=\frac{\partial \Omega_{\mathrm{Q}}}{\partial d_{i}}.
\end{equation}
From the conditions for electromagnetic charge neutrality and color charge neutrality, we have
\begin{equation}
n_{j}=-\frac{\partial \Omega_{\mathrm{Q}}}{\partial \mu_{j}}=0,
\end{equation}
where $j = 3,8, Q$. 
The baryon number density $n_{B}$ is determined as
\begin{equation}
n_{q}=-\frac{\partial \Omega_{\mathrm{Q}}}{\partial \mu_{q}},
\end{equation}
where $\mu_{q}$ is $1/3$ of the baryon number chemical potential. After determined all the values, we obtain the pressure as
\begin{equation}
P_{\mathrm{Q}}=-\Omega_{\mathrm{Q}}.
\end{equation}

\subsection{FIRST-ORDER PHASE TRANSITION}

In this section, we construct the EOS for NS matter incorporating a first-order phase transition\cite{Lenzi:2012xz,Benic:2014jia}.   The hadronic phase is described by the PDM, which includes the parameter $m_0$, representing the chiral invariant mass. The quark phase is modeled using an NJL-type quark model with two key parameters: $H$, characterizing the strength of attractive diquark correlations, and $g_V$, denoting the strength of vector repulsion. By adjusting the values of these parameters, we can control the stiffness of the quark matter EOS and, consequently, the onset and properties of the phase transition.

\begin{figure}[htbp]\centering
\includegraphics[width=1\hsize]{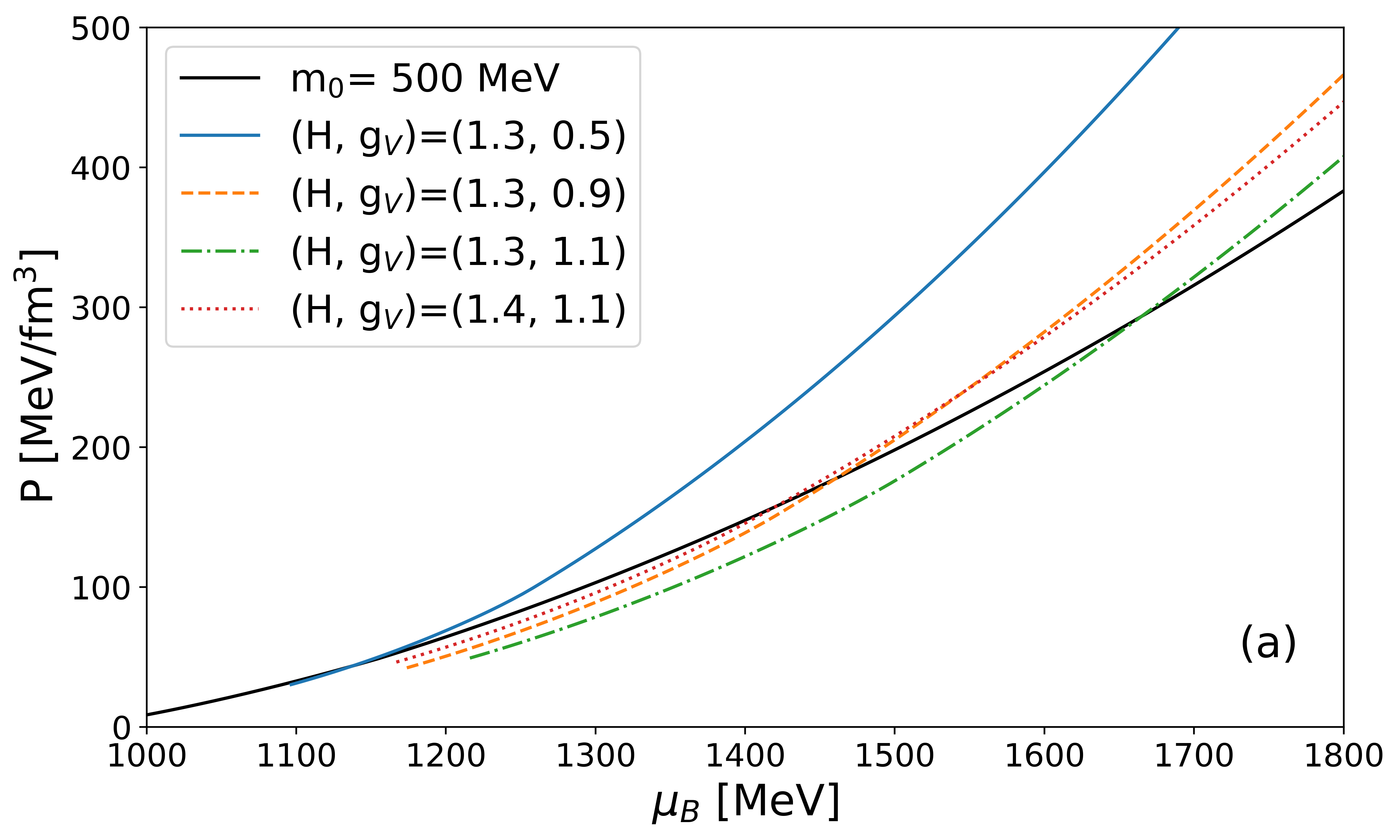}
\includegraphics[width=1\hsize]{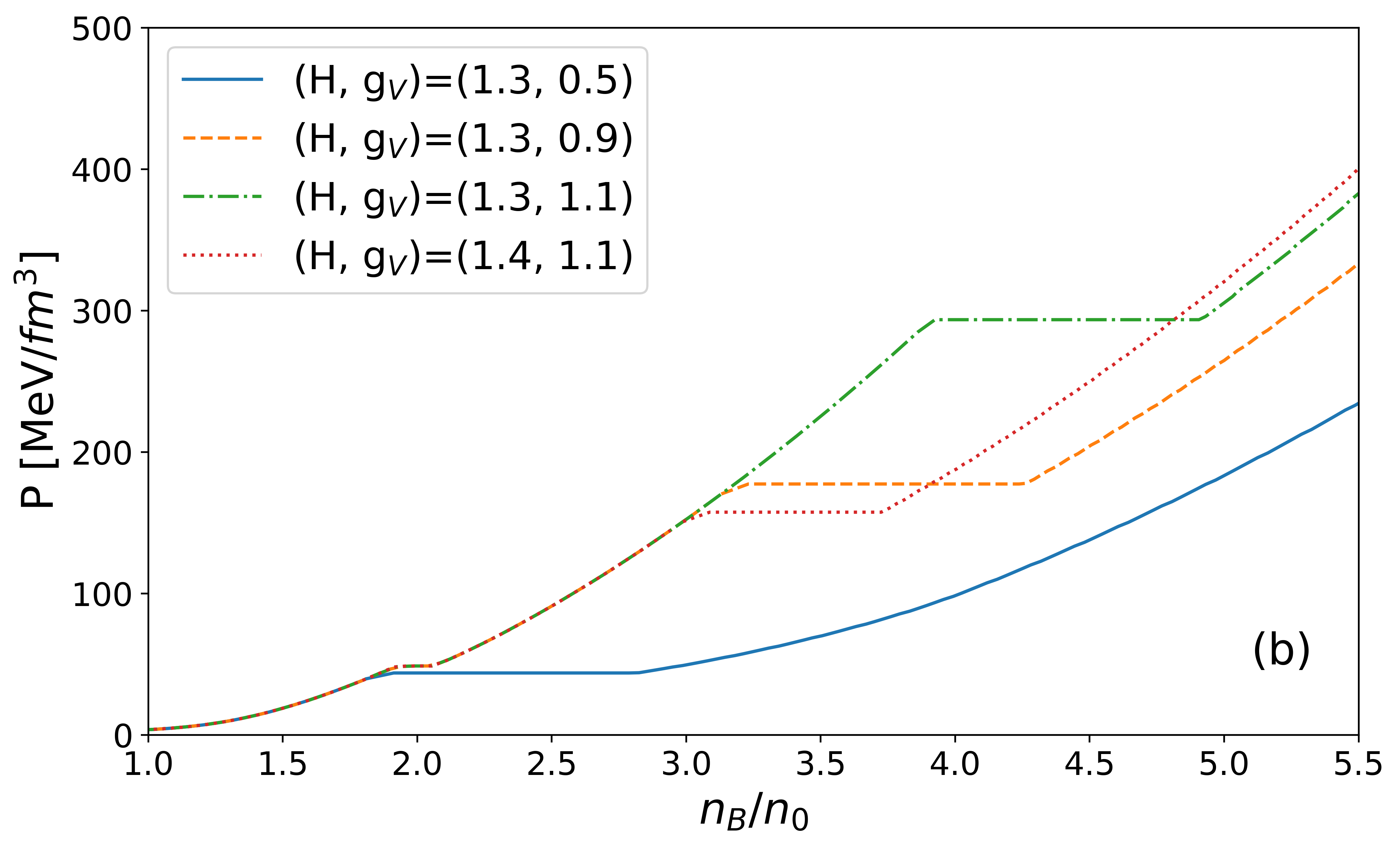}
\caption{(a)Pressure as a function of baryon chemical potential for fixed value of $m_0$ in the PDM and several choices of ($H, g_{V}$) combination in the NJL-type quark model (upper panel). (b)Pressure as a function of baryon number density in unit of normal nuclear density $n_0$ constructed from (a). }
\label{p_mu_500}
\end{figure}

To construct a hybrid EOS with a first-order phase transition, we first fix the value of $m_0$ in the PDM. As an example, we set $m_0 = 500$ MeV, which corresponds to a relatively stiff hadronic EOS. We then explore various combinations of $(H, g_V)$ in the NJL-type quark model to find an intersection point between the hadronic and quark matter EOSs, as shown in Fig. \ref{p_mu_500}(a). The intersection point indicates the baryon chemical potential at which the phase transition occurs, ensuring that the pressure and chemical potential of both phases are equal, satisfying the Maxwell construction~\cite{Zdunik:2012dj,Alford:2013aca} for a first-order phase transition. The resulting hybrid EOS is presented in Fig.~\ref{p_mu_500}(b). If the transition density from hadronic matter to quark matter is larger than density where chiral phase transition---the degeneracy of the positive and negative parity baryons---can happen, then two distinct plateaus are observed in the EOS for $(H, g_V) = (1.3, 0.5), (1.3, 0.9), (1.3, 1.1)$. The first plateau, located around $2n_0$, corresponds to the emergence of the negative parity baryon $N(1535)$, a characteristic feature of the PDM. This transition is associated with a softening of the EOS due to the partial restoration of chiral symmetry. The second plateau represents the first-order phase transition from hadronic matter to quark matter. The onset of this transition is determined by the chosen values of $(H, g_V)$ in the NJL-type quark model. During the phase transition, the pressure remains constant while the baryon density experiences a discontinuous jump, indicating a coexistence region between hadronic and quark matter phases. On the other hand, if the quark-hadron transition density is smaller than the chiral phase transition, there is only one plateau as the blue curve $(H, g_V) = (1.3, 0.5)$ shows.

The presence of these two transitions in the hybrid EOS has significant implications for the structure and properties of neutron stars. The softening of the EOS due to the transitions can impact the maximum mass and radius of neutron stars, as well as their tidal deformability and other observable properties. By systematically varying the parameters $m_0$ in the PDM and $(H, g_V)$ in the NJL-type quark model, we can explore a range of possible hybrid EOSs with first-order phase transition and assess their compatibility with observational constraints from NSs.

\section{Neutron star properties}
\label{sec-NS}
In this section, we study the NS properties by using the EOSs constructed in the previous section. 
%
%
%
By solving the Tolman-Oppenheimer-Volkoff (TOV) equation for spherically symmetric and static stars, we obtain the NS mass-radius (M-R) relation.

\begin{figure}[htbp]\centering
\includegraphics[width=1\hsize]{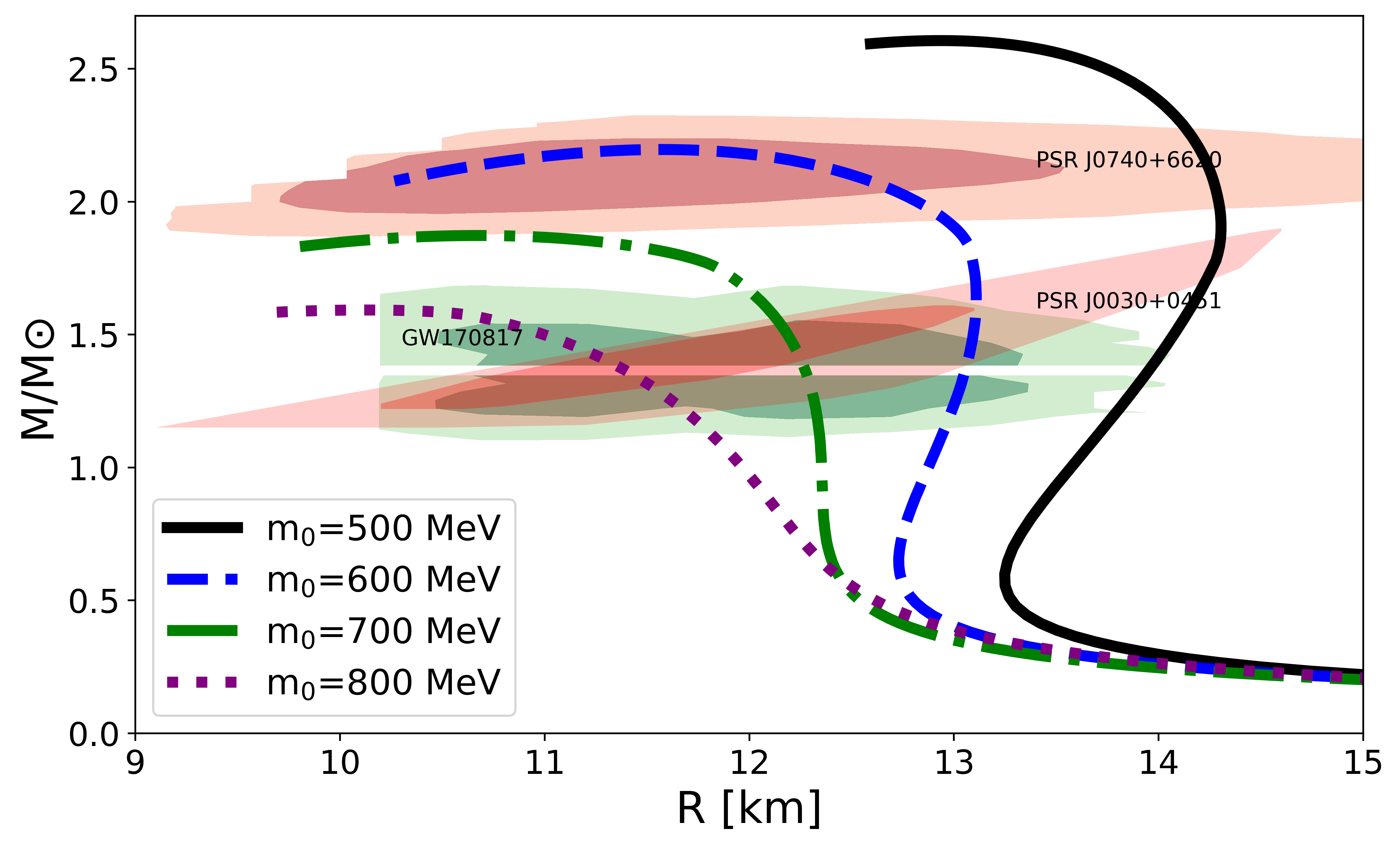}
\caption{Mass-radius relation in PDM for different values of $m_0$. }
\label{purePDM}
\end{figure}

We first show the M-R relation in pure hadronic PDM in Fig.~\ref{purePDM}. The green and red contours represent the NS observational constraints from LIGO-VIRGO~\cite{PhysRevLett.119.161101,LIGOScientific:2017ync,LIGOScientific:2018cki} and the Neutron Star Interior Composition Explorer (NICER)~\cite{Miller:2021qha, Riley:2021pdl, Vinciguerra:2023qxq}, respectively. We observe that for $m_0$ = 700 and 800 MeV, the maximum mass of the NS is smaller than 2$M_{\odot}$  constraint. Since the first-order phase transition soften the EOS and lower the maximum mass of the NS, it is impossible to realize the NS with $M/M_{\odot} \geq 2$ for $m_0 \geq 700$ MeV. Consequently, we focus on the cases with $m_0$ = 500 and 600 MeV in the subsequent analysis. 

\begin{figure*}[htbp]
	\centering
	\subfloat[]{
		\includegraphics[width=0.9\textwidth]{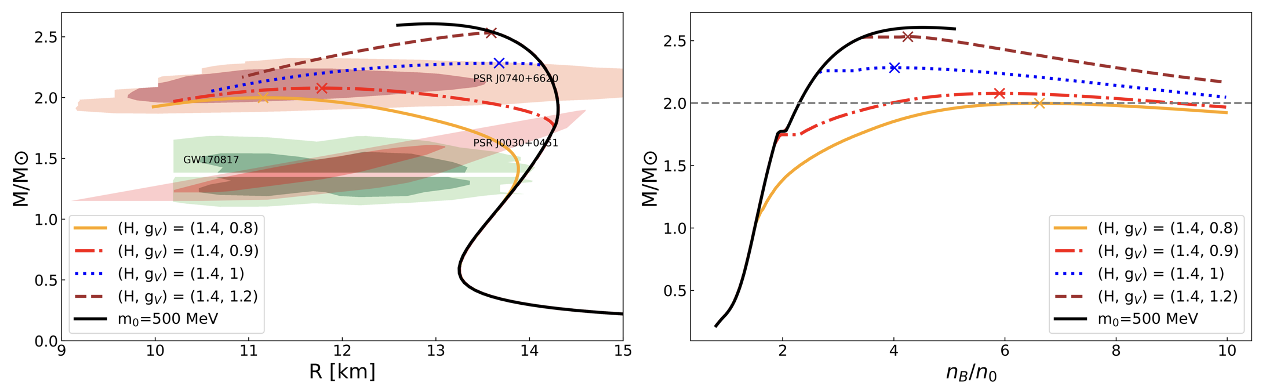}
		\label{fig:first}
	}
	\hfill
	\subfloat[]{
		\includegraphics[width=0.9\textwidth]{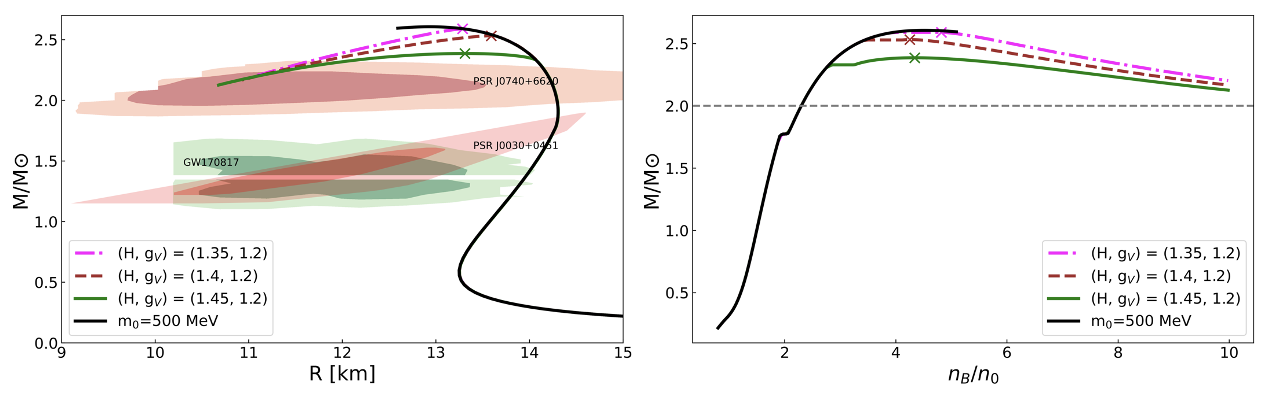}
		\label{fig:second}
	}
	\caption{Mass-radius relation for $m_0=500$ MeV combined with different choices of NJL parameters and corresponding mass-center density relation. In (a), we fix the $H$ value and change the $g_{V}$ parameter while in (b) we fix the $g_{V}$ parameter while change the $H$.}
	\label{fig:500}
\end{figure*}

\begin{figure*}[htbp]\centering
	\includegraphics[width=0.9\hsize]{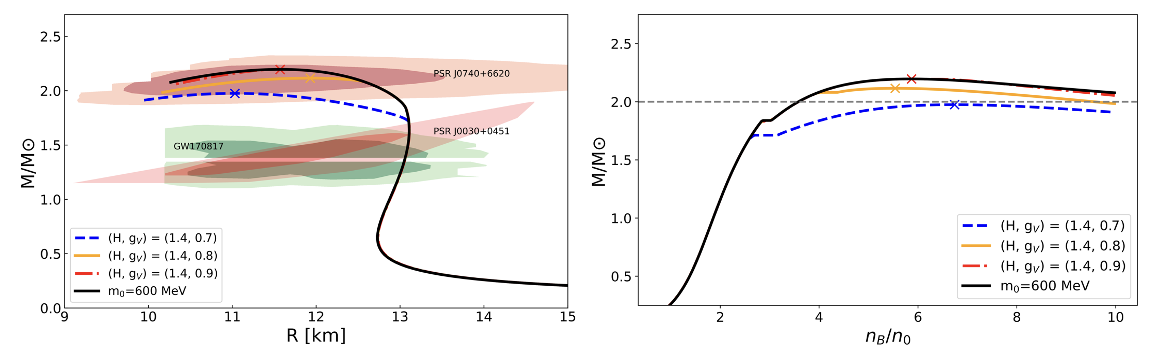}
	\caption{Mass-radius relation and mass-center density relation for $m_0=600$ MeV with different NJL parameters.}
	\label{fig:600}
\end{figure*}

Next, we examine the impact of the first-order phase transition on the M-R relation by combining the PDM with the NJL-type model. We show some typical example in Fig.~\ref{fig:500} and Fig.~\ref{fig:600}, with the black line representing the pure PDM and different colors corresponding to various NJL parameter combinations $(H, g_{V})$. For the purpose to investigate the effects of parameters $H$ and $g_V$, we change one values with keeping the others fixed, as shown in Fig.~\ref{fig:500}(a) and Fig.~\ref{fig:500}(b). The cross marks indicate the points of maximum mass for each combination.

From Fig.~\ref{fig:500}, one can conclude that the increasing of $g_{V}$ or the decreasing of $H$ increases the onset density of the first-order phase transition. When the phase transition occurs at a very low density, as indicated by the orange line in Fig. \ref{fig:500}, the magnitude of chiral symmetry breaking is suppressed. As the onset density of the first-order phase transition is pushed to higher densities by increasing $g_V$ or decreasing $H$, the chiral restoration around 2$n_0$ for $m_0 = 500$ MeV begins to manifest. In the figure, the orange, red, and blue lines represent stable NS configurations, with the maximum mass $\sim 2m_\odot$ appearing after the first-order phase transition. However, the brown line and the magenta line in Fig. \ref{fig:500}(b) exhibit a significant reduction in radius and mass following the phase transition, implying an NS with a unstable quark core since the ${\rm d} M / {\rm d} n_B < 0$ after the phase transition~\cite{Alford:2017qgh, Glendenning:1998ag}. 

We also show the case for $m_0=600$ MeV in Fig.~\ref{fig:600} with some typical choices of the NJL parameters. In this case, the onset of $N(1535)$ point is shifted to around $3n_0$ and a softer quark EOS is required to form a stable NS compared with the case in $m_0=500$ MeV. In the two figures in Fig.~\ref{fig:600}, we find that the onset density of the phase transition, governed by the interplay between hadronic and quark matter parameters, significantly influences the stability of the NS configuration. A phase transition occurring at lower densities allows for stable NS configurations, while a phase transition at excessively high densities can lead to an unstable NS, leads to ${\rm d} M / {\rm d} n_B < 0$ after the phase transition. This is because if the onset density is very high, only NSs with very high central densities will undergo this phase transition. These stars are typically very massive and close to the TOV limit (the maximum mass limit for a stable neutron star). At such high densities, the EoS change due to the phase transition can make the core more compressible or softer, which can destabilize the star. This destabilization may lead to the collapse of the NS star into a black hole if the pressure support is insufficient to counteract the gravitational pull.

\begin{figure*}[htbp]
    \centering
    \subfloat[]{
        \includegraphics[width=0.47\textwidth]{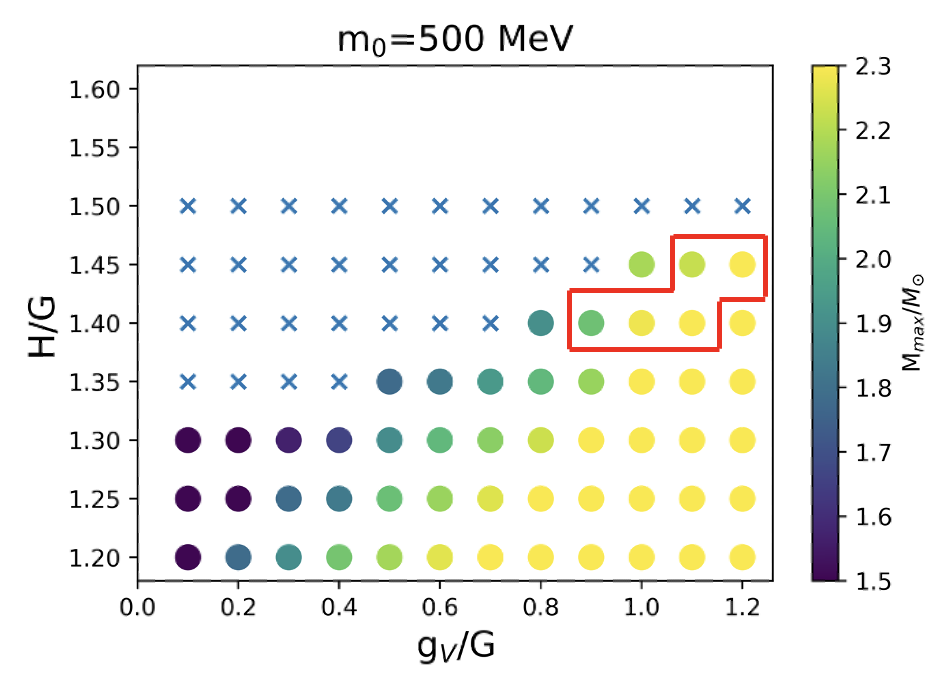}
        \label{fig:first}
    }
    \hfill
    \subfloat[]{
        \includegraphics[width=0.47\textwidth]{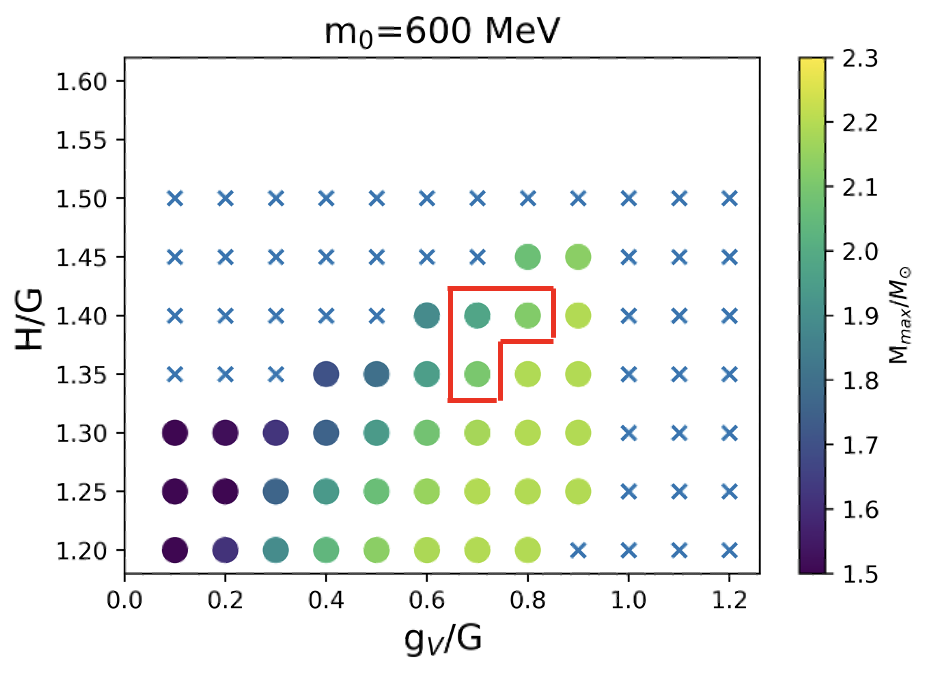}
        \label{fig:second}
    }
        \subfloat[]{
        \includegraphics[width=0.47\textwidth]{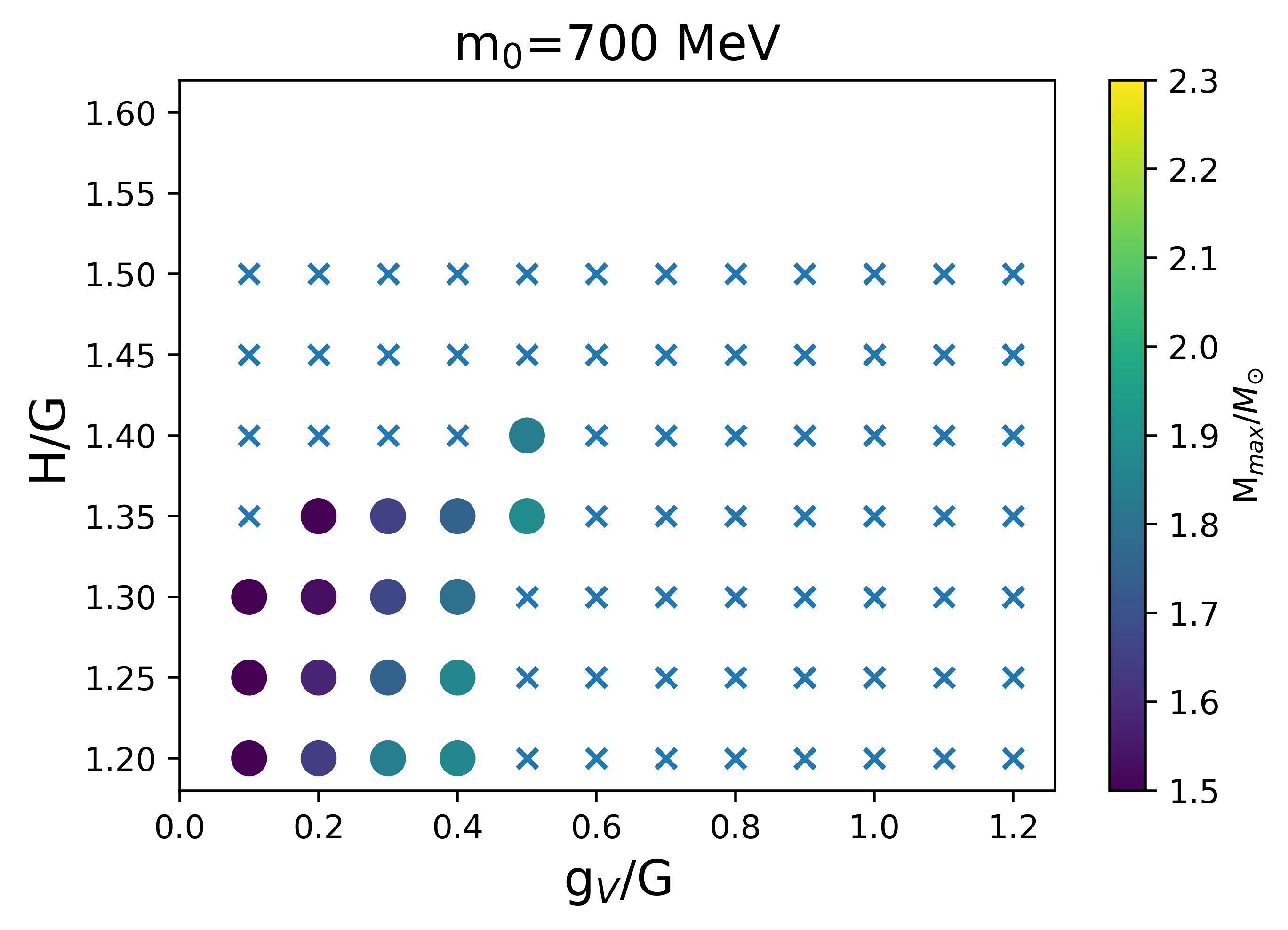}
        \label{fig:second}
    }
    \caption{Possible $(H, g_{V})$ combination for $m_0= 500, 600, 700$ MeV. Cross mark indicates that the EOS constructed from the combination of ($H, g_V$) do not have intersect point with EOS constructed from PDM. Circle indicates that the combination is allowed. The color shows the maximum mass of NS obtained from the corresponding parameters, as indicated by a vertical bar at the right side of each figure. The part surrounded by red lines represents the combination can satisfy the maximum mass constraints and the stable condition at the same time.}
    \label{contour}
\end{figure*}

Finally, by fine-tuning the NJL parameters, we can identify the suitable parameter space and constrain the onset density of the first-order phase transition. In Fig. \ref{contour}, we show all possible combinations of $(H, g_{V})$ for $m_0$ = 500, 600 and 700 MeV. The color of each point indicates the maximum mass of the NS obtained from the corresponding parameters, as shown by the vertical bar on the right side of each figure. A cross mark indicates that the EOS constructed from the combination of $(H, g_{V})$ does not have an intersection point with the EOS constructed from the PDM, while a colored indicates that the combination is allowed. For $m_0$=700 MeV, all the combination of NJL parameters lead to a NS maximum mass smaller than 2$M_{\odot}$, leading to the conclusion that the $m_0$ should be smaller than 700 MeV while considering the first-order phase transition.

After considering NS with a stable quark core, we further constrain the parameter space to the region surrounded by red lines which represents the combinations that can simultaneously satisfy the maximum mass constraints and the stability condition. Using these parameters, we can determine the possible range of densities at which the first-order phase transition occurs.

\begin{figure}[htbp]\centering
\includegraphics[width=1\hsize]{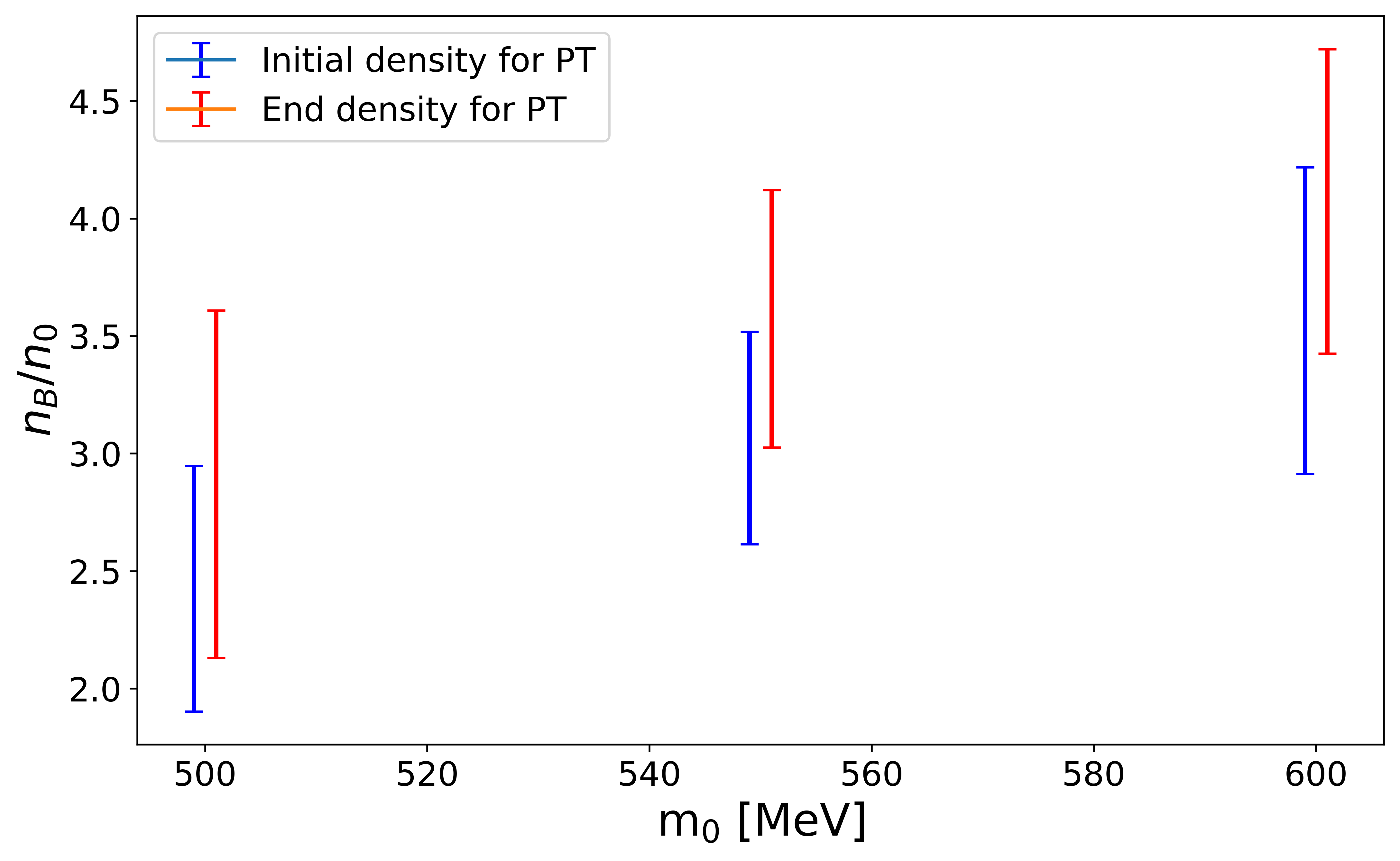}
\caption{Density range of possible first-order phase transition as a function of $m_0$. }
\label{PT_density}
\end{figure}

We choose $m_0$ to be 500, 550, and 600 MeV and examine the density range of the first-order phase transition as shown in Fig. \ref{PT_density}. The blue line in the figure represents the density range at which the phase transition begins, while the red line represents the density range where the phase transition ends. As the value of $m_0$ increases, we find that both the onset density and the ending density of the phase transition increase.
For $m_0$ = 500 MeV, the phase transition occurs between $1.9n_0$ and $2.95n_0$ and ends between $2.1n_0$ and $3.6n_0$. When $m_0$ = 600 MeV, the phase transition takes place between $2.9n_0$ and $4.1n_0$ and concludes between $3.4n_0$ and $4.6n_0$.

\section{Summary and Discussion}
\label{sec-conclusion}

In this paper, we have investigated the possibility and impacts of a first-order phase transition from hadronic matter to quark matter in NSs using the PDM for the hadronic phase and the NJL-type model for the quark phase. By constructing hybrid EOSs and analyzing the M-R relations, we have explored the role of key parameters in both models, such as the chiral invariant mass ($m_0$) in the PDM and the attractive diquark correlation strength ($H$) and vector repulsion strength ($g_V$) in the NJL-type quark model.

Our findings show that the onset density of the first-order phase transition is strongly influenced by the interplay between hadronic and quark matter parameters. A smaller chiral invariant mass $m_0$ in the PDM leads to an earlier onset of the chiral phase transition associated with the emergence of the negative parity baryon. The stability of the NS configuration is sensitive to the onset density of the first-order phase transition, with transitions occurring at lower densities allowing for stable configurations, while transitions at excessively high densities can lead to NS with an unstable quark core.
By fine-tuning the parameters $(H, g_V)$ and considering the stability of the NS, we have identified suitable parameter spaces and constrained the onset density of the first-order phase transition. For $m_0$ = 500 MeV, the phase transition occurs between 1.9$n_0$ and 2.95$n_0$ and ends between 2.1$n_0$ and 3.6$n_0$. For $m_0$ = 600 MeV, the phase transition occurs between 2.9$n_0$ and 4.1$n_0$ and ends between 3.4$n_0$ and 4.6$n_0$.

\begin{figure*}[htbp]\centering
	\includegraphics[width=0.9\hsize]{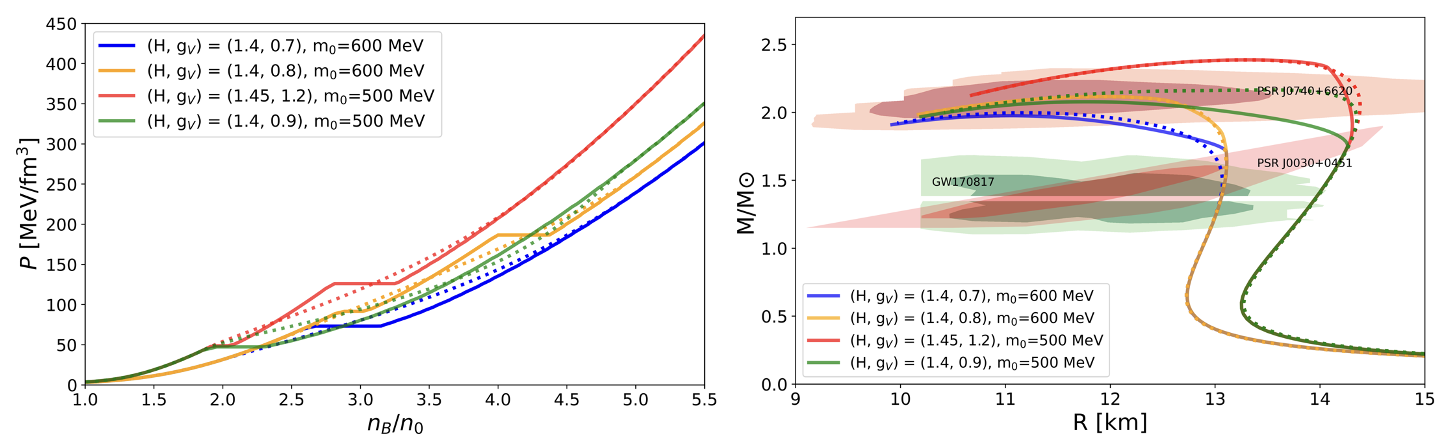}
	\caption{Comparison of the 1st order phase transition with the crossover transition. Left panel shows pressure as the function of density and right panel shows the corresponding $M$-$R$ curve for $m_0=500, 600$ MeV with different choices of NJL parameters. Solid curves are for the 1st order phase transition and the dotted curves are for the crossover transition.}
	\label{fig:crossover}
\end{figure*}

Finally, we compare cases assuming a crossover transition with those involving a first-order phase transition, as illustrated in Fig. \ref{fig:crossover}. For each $m_0$ value, we select NJL parameters from the  region surrounded by red lines in Fig.~\ref{contour}  that yield both the largest and smallest maximum masses. Solid lines represent cases with first-order phase transitions, while dotted lines of the same color indicate cases assuming crossover transitions. Our analysis reveals several key differences between these two scenarios. We observe that the crossover assumption leads to a smooth change in the $M$-$R$ curve, while the first-order phase transition results in a more abrupt change, often characterized by a "kink" in the M-R relation. Furthermore, the maximum neutron star mass in the first-order phase transition scenario is typically smaller than in the crossover case, which can be attributed to the sudden softening of the equation of state during a first-order transition.

\section*{Acknowledgments}

The work of  B.G., and M.H. are supported in part by JSPS KAKENHI Grant Nos.~ 23H05439 and 24K07045.
B.G. is also supported by JST SPRING, Grant No. JPMJSP2125.  B.G. would like to take this opportunity to thank the
“Interdisciplinary Frontier Next-Generation Researcher Program of the Tokai
Higher Education and Research System.”
The work of Y.~L. M. was supported in part by the National Key R\&D Program of China under Grant No. 2021YFC2202900 and the National Science Foundation of China (NSFC) under Grant No. 12347103 and No. 11875147.

\bibliography{ref_3fPDM_2022}

\end{document}